\documentclass[12pt]{iopart}
\usepackage{graphicx}
\usepackage{subfigure}
\begin{document}

\title [$\pi$, K, p spectra with ALICE in pp collisions at $\sqrt{s} =$ 0.9 and 7 TeV]{Measurement of $\pi$, K, p  transverse momentum spectra with ALICE in proton-proton collisions at $\sqrt{s} =$ 0.9  and 7 TeV}

\author{Marek Chojnacki for the ALICE collaboration }

\address{Utrecht University, Princetonplein 5, 3584 CC Utrecht, Netherlands}
\ead{Marek.Chojnacki@cern.ch}
\begin{abstract}
Results of the measurement of the $\pi$, K, p transverse momentum ($p_{\mathrm{t}}$) spectra at mid-rapidity in proton-proton collisions at $\sqrt{s} = 7$ TeV are presented.   
Particle identification was performed using the energy loss signal in the Inner Tracking System (ITS) and the Time Projection Chamber (TPC), while information
from the Time-of-Flight (TOF) detector was used to identify particles at higher transverse momentum.  
From the spectra at $\sqrt{s} = 7$ TeV the mean transverse momentum ($\langle p_{\mathrm{t}}\rangle$) and particle ratios were extracted and compared to results obtained for collisions at  $\sqrt{s} = 0.9$ TeV and lower energies. 
\end{abstract}
\section{Introduction}
The production of particles in ultra-relativistic proton-proton collisions with a transverse momentum ($p_{\mathrm{t}}$) 
close to 1 GeV/\textit{c} at mid-rapidity, can't be uniquely described by perturbative Quantum Chromodynamics. 
It has to be measured, to provide an input for theoretical developments. 
\newline
In 2010 the LHC produced first proton-proton collisions  at $\sqrt{s} = 7$ TeV, entering a new energy regime for such studies.
Using its particle identification capabilities  \cite{alex}, the ALICE experiment was capable  of measuring the  pion, kaon and proton $p_{\mathrm{t}}$-spectra.
The same measurements for  $\sqrt{s} = 0.9$ TeV proton-proton collisions collected in December 2009  were presented in \cite{900GeV}, and they are used for comparison with the 7 TeV data.
\section{Methods of measurement}
Four methods of extracting raw yields of $\pi$, K, p were used: 
\begin{itemize}
\item ITS - the measurement was done using standalone ITS tracks\footnote{Standalone ITS tracks are reconstructed using only information from the Inner Tracking System}. 
The extraction of the raw yield was performed by selecting particles with the measured energy loss signal in four ITS layers close to the expected one for each particle species.
\item ITSTPC - the measurement was done using global tracks\footnote{Global tracks are reconstructed using information from the ITS and the TPC.}. 
The extraction of the raw yield was performed by fitting an analytical function to the distribution of the energy loss signal in four ITS layers  in a given $p_{\mathrm{t}}$ bin.  
\item TPCTOF - the measurement was done using global tracks. 
The extraction of the raw yield was performed by selecting particles with the measured energy loss signal in the TPC (measured time of flight in the TOF for higher $p_{\mathrm{t}}$ ) close to the expected one for each particle type. 
\item TOF - the measurement was done using global tracks. 
The extraction of the raw yield was performed by fitting an analytical function to the  time of flight distribution in a given $p_{\mathrm{t}}$ bin. 
\end{itemize}
\begin{figure}
\centering
\subfigure{\includegraphics[width=12pc]{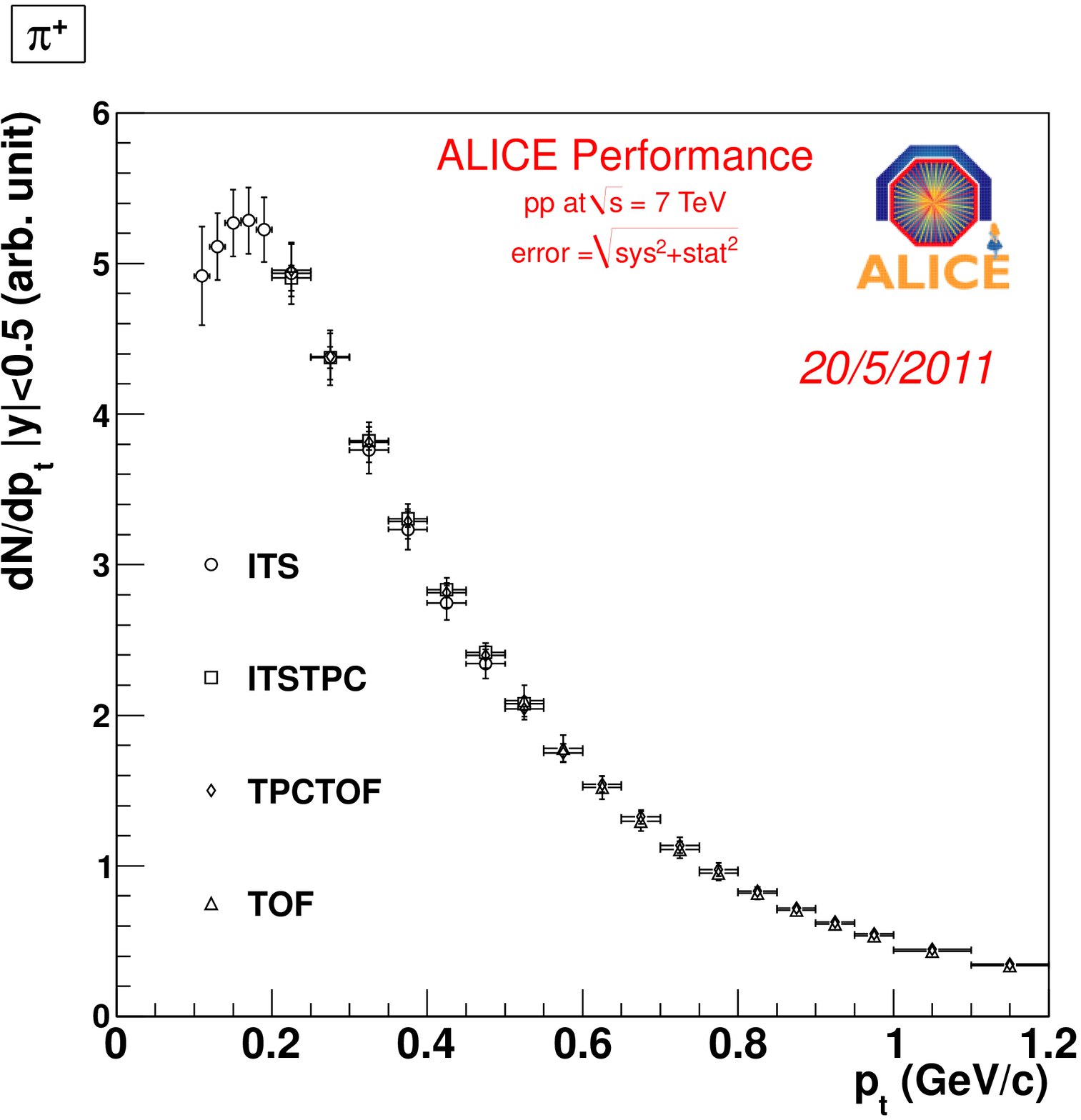}}
\subfigure{\includegraphics[width=12pc]{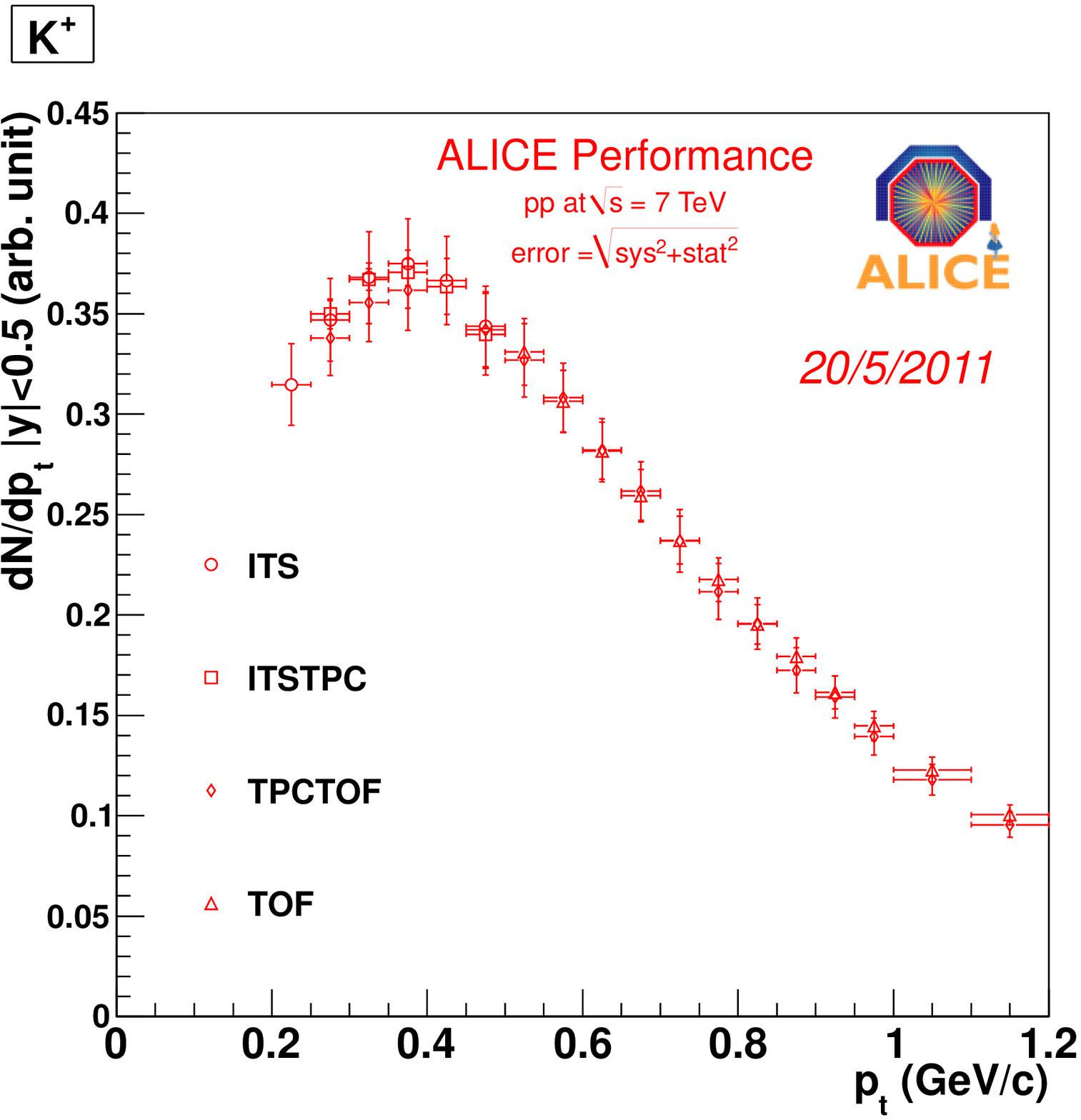}}
\subfigure{\includegraphics[width=12pc]{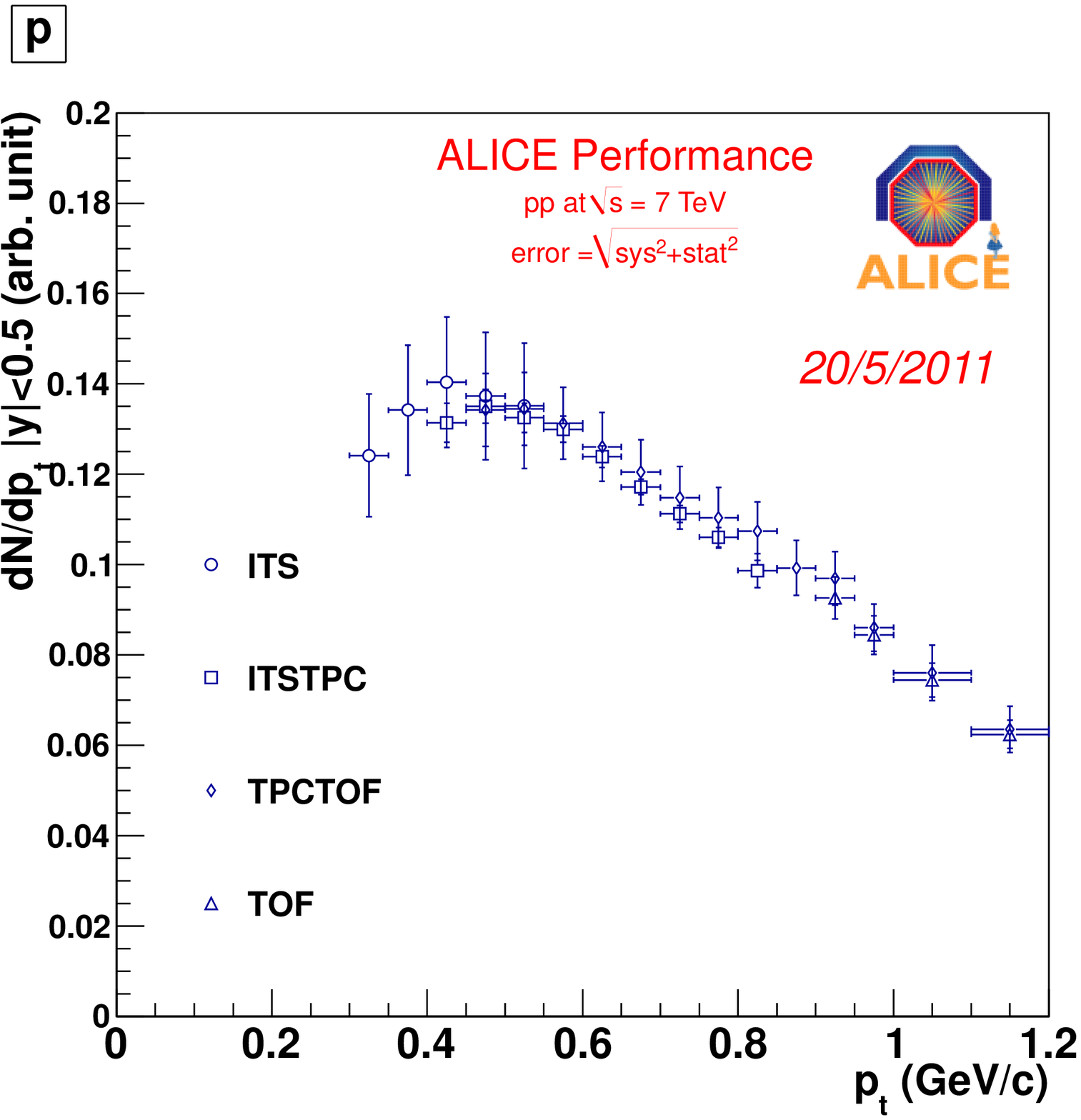}}
\caption{Corrected spectra for positively charged particles measured by 4 methods in proton-proton collisions at $\sqrt{s}=7$ TeV.
A $p_{\mathrm{t}}$-region where all methods can be applied is shown.
The total uncertainties (systematic and statistical added in a quadrature) are plotted.} 
\label{fig:4Methods}
\end{figure}
Each method was used in a different $p_{\mathrm{t}}$ range.
The raw yields were corrected for the tracking inefficiency using MC simulations of the performance of the ALICE detector.  
PYTHIA and  PHOJET event generators were used together with GEANT3 transport code.  
The corrections for secondary protons and antiprotons (products of weak decays and interactions with detector material)
were determined by comparing the distributions of the transverse impact parameter in data and simulations. 
For the TPCTOF method the same procedure was used to extract the pion contamination. 
For other methods this correction was based on simulations.
All corrected spectra were calculated in the rapidity region $|y|<0.5$.
The rapidity selection was made using the relevant mass hypothesis for each particle types, detail description of that method can be found in \cite{900GeV}.
\newline 
Figure \ref{fig:4Methods} demonstrates  the agreement between the four different methods in \mbox{$p_{\mathrm{t}}$=(0.0-1.2) GeV/\textit{c}} region. 
The agreement  is well within the bounds of the total uncertainties (systematic and statistical added in a quadrature).
The spectra from the four methods were combined by averaging, using the total uncertainties as weights. 
The combined spectra were normalized to the number of inelastic events \cite{ken,Martin} with the uncertainty of 8.3\%.
During the analysis data sample which contains 8 million minimum-bias events data was used.          
\section{Results and Discussion}
\begin{figure}[!h]
\centering
\subfigure{\includegraphics[width=.49\textwidth]{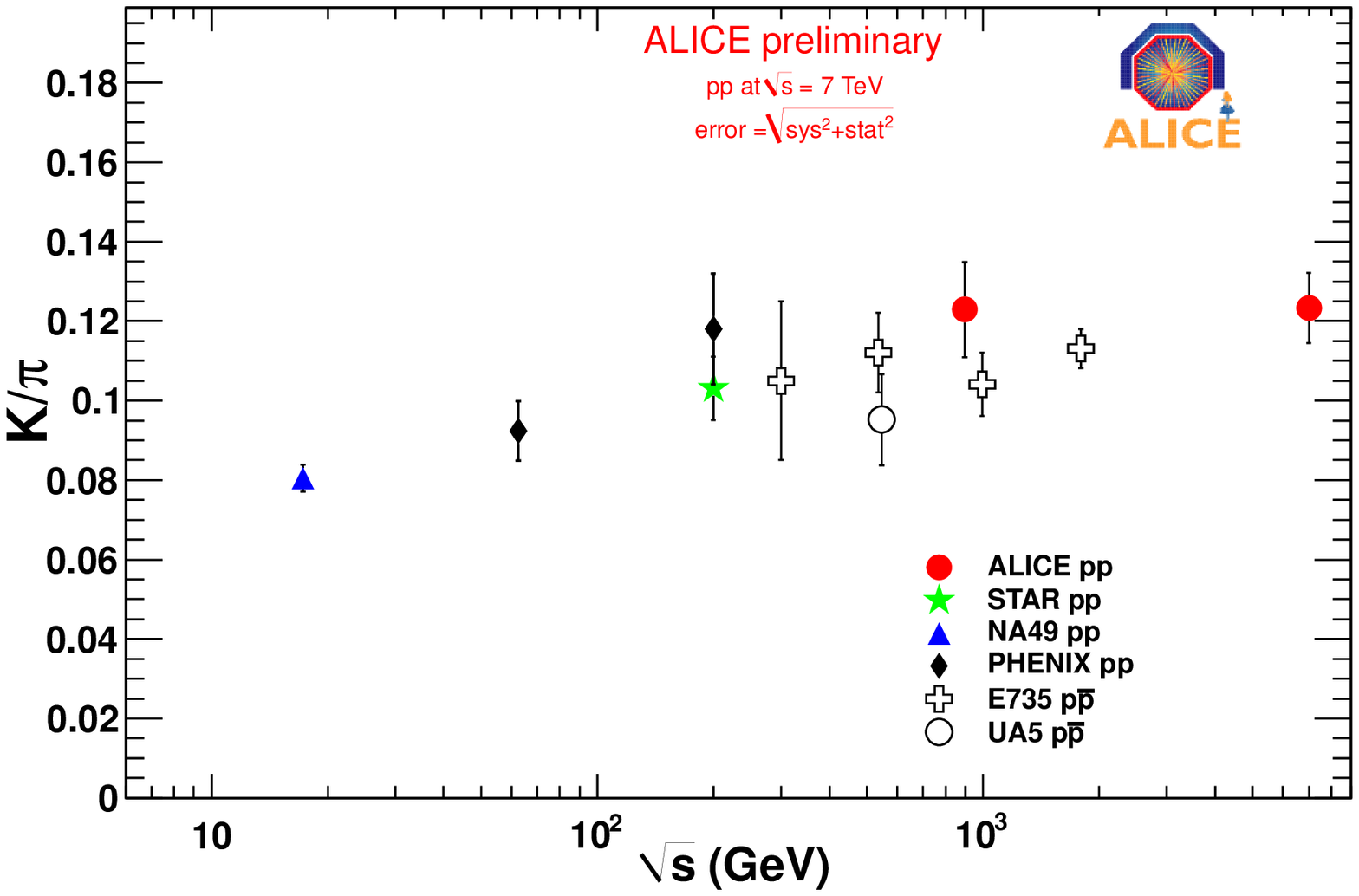}}
\subfigure{\includegraphics[width=.49\textwidth]{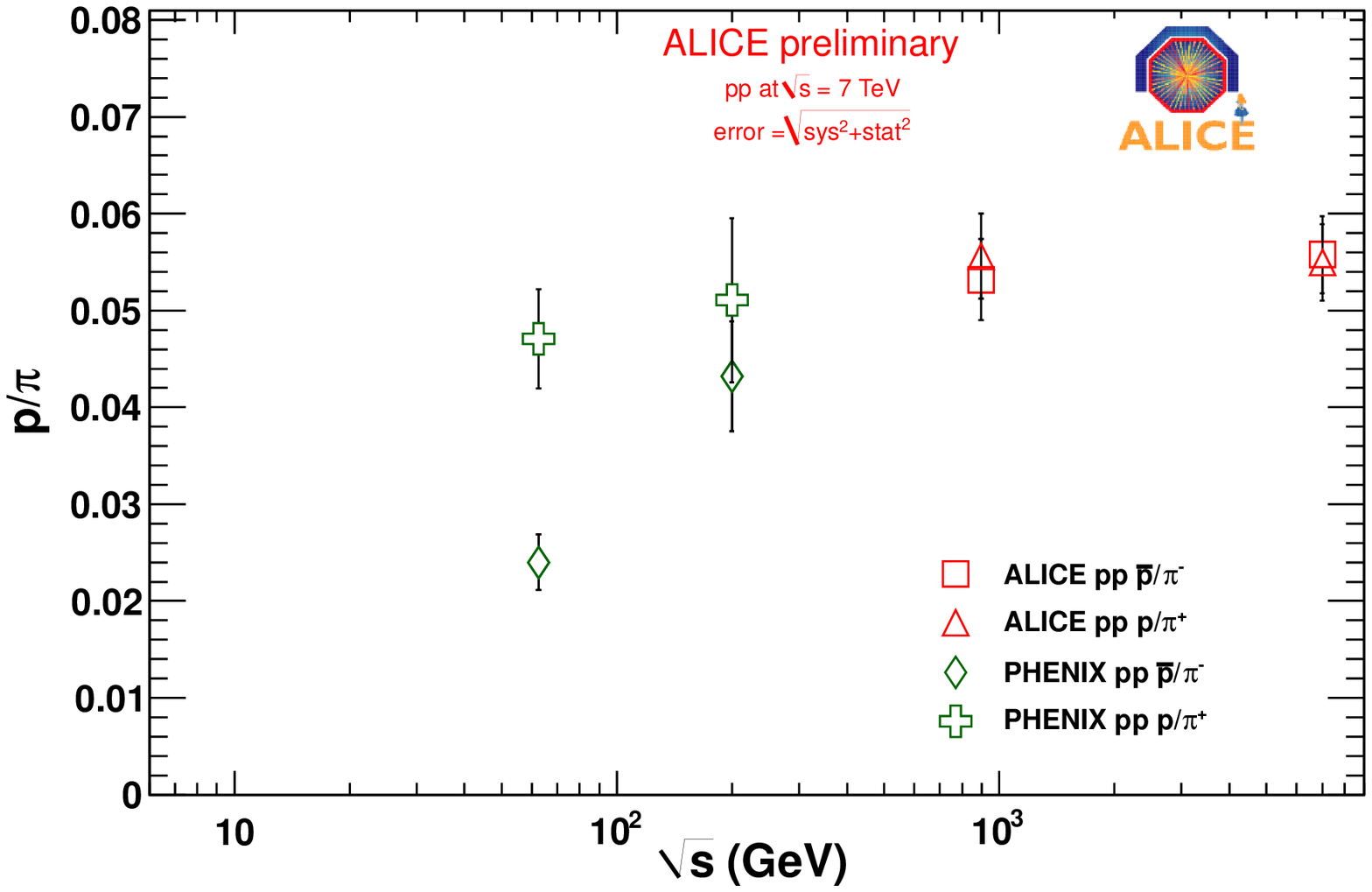}}
\caption{Evolution of the integrated yields ratios $(K^{+}+K^{-})/(\pi^{+}+\pi^{-})$, $K^{0}/\pi$ (left) and $p/\pi^{+}$, $\bar{p}/\pi^{-}$ (right) as function of collision energy in proton-proton and proton-antiproton collisions \cite{900GeV,phenix}.} 
\label{fig:ratios}
\end{figure}

The combined and normalized spectra were fitted using the  L\'{e}vy function:
\begin{equation}
\frac{\rmd^{2} N}{\rmd p_{\mathrm{t}}\rmd y}=p_{\mathrm{t}}\frac{\rmd N}{\rmd y}\frac{(n-1)(n-2)}{nC(nC+m_{0}(n-2))}\left(1+\frac{m_{\mathrm{t}}-m_{0}}{nC}\right)^{-n} 
\label{eq: Levy}
\end{equation}
\begin{figure}
\begin{center}
\includegraphics[width=24pc]{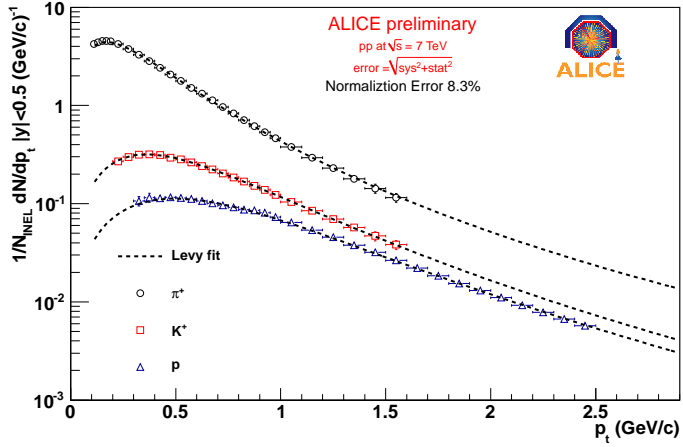}
\caption{Combined  $p_{\mathrm{t}}$-spectra for positively charged particles in proton-proton collisions at $\sqrt{s}=7$ TeV. 
Lines are the fits of the  L\'{e}vy function (equation \ref{eq: Levy}).
Beside total uncertainties there is additional 8.3\% uncertainty due to the normalization.}   
\label{fig:fits}
\end{center}
\end{figure}
\begin{figure}
\begin{center}
\includegraphics[width=24pc]{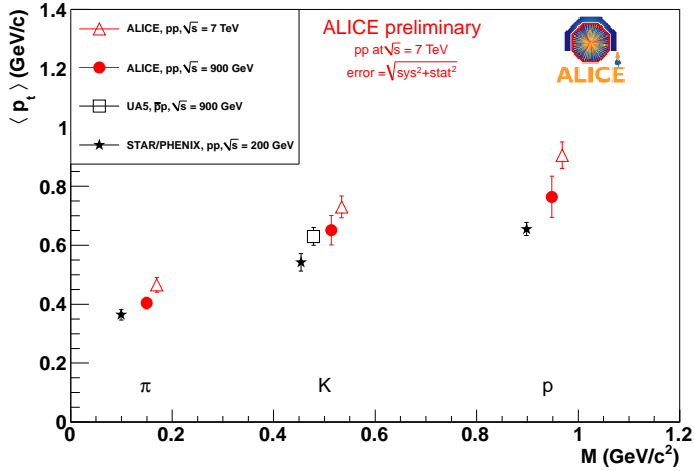}
\caption{ $\langle p_{\mathrm{t}}\rangle$ as function of particle  mass. 
Results are given for different proton-proton and (proton-antiproton) collision energies \cite{900GeV}.}   
\label{fig:meanpt}
\end{center}
\end{figure}
The fits for the positively charged particle spectra are shown in Figure \ref{fig:fits}.
The agreement between the data points and the  fits is on the level of a few percent. 
The  L\'{e}vy function  also describes well the spectra measured by the ALICE experiment at $\sqrt{s} = 0.9$ TeV proton-proton collisions \cite{900GeV}.
The $\langle p_{\mathrm{t}}\rangle$  and  the integrated yield ($\frac{\rmd N}{\rmd y}$) for all particles were calculated from the  L\'{e}vy function fits. 
To evaluate the systematic uncertainties different spectra extrapolation scenarios to the region not covered by the measurement were tested.  
The  $\langle p_{\mathrm{t}}\rangle$ value for each particle type was taken the average $\langle p_{\mathrm{t}}\rangle$ value for positive and negative particles. 
The difference between $\langle p_{\mathrm{t}}\rangle$ measured for the positive and negative particles of that same species is much smaller than the extrapolation uncertainty.
\newline
Figure \ref{fig:meanpt} shows   $\langle p_{\mathrm{t}}\rangle$ for $\pi$, K, p for different collisions energies. 
$\langle p_{\mathrm{t}}\rangle$  increases with energy and with the particle mass. 
Using the extrapolated   values of the ($\frac{\rmd N}{\rmd y}$)  ratios $(K^{+}+K^{-})/(\pi^{+}+\pi^{-})$, $p/\pi^{+}$ and $\bar{p}/\pi^{-}$  were calculated. 
They were compared to ratios measured at lower energies, as shown in Fig. \ref{fig:ratios}. 
No strong energy dependence is observed between proton-proton collisions at  $\sqrt{s} =$ 0.9 and 7 TeV.

\section{Summary}
As the energy in proton-proton collision increases from 0.9 to 7 TeV,
an increase of the $\langle p_{\mathrm{t}}\rangle$ for pions, kaons and protons is observed, 
no significant change in relative particle yields ($(K^{+}+K^{-})/(\pi^{+}+\pi^{-})$, $p/\pi^{+}$, $\bar{p}/\pi^{-}$) is found.


\section*{References}
\begin{thebibliography}{10}
\bibitem{alex} A. Kalweit, these proceedings
\bibitem{900GeV} K. Aamodt, \textit{et al.} [ALICE Collaboration], Eur.Phys.J.C 71(6): 1655, 2011

\bibitem{ken} K. Oyama, these proceedings
\bibitem{Martin} M. Poghosyan, these proceedings
\bibitem{phenix} A. Adare, \textit{et al.} [PHENIX Collaboration], Phys. Rev. C 83, 064903 (2011) 


\end{thebibliography}
\end{document}